\documentclass{article}
\usepackage{iclr2025_conference,times}
\usepackage{hyperref}
\usepackage{url}
\usepackage{graphicx}
\usepackage{amsmath}
\usepackage{booktabs}
\usepackage{caption}
\usepackage{subcaption}
\usepackage{float}
\usepackage{enumitem}

\iclrfinalcopy

\title{Peeking Behind Closed Doors: Risks of LLM Evaluation by Private Data Curators}

\author{
    Hritik Bansal\thanks{Equal contribution} \\
    UCLA \\
    \texttt{\href{https://sites.google.com/view/hbansal}{hbansal@g.ucla.edu}} \\
    \And
    Pratyush Maini\footnotemark[1] \\
    CMU \& DatologyAI \\
    \texttt{\href{http://pratyushmaini.github.io/}{pratyushmaini@cmu.edu}}
}

\begin{document}
\maketitle

\begin{abstract}
The rapid advancement in building large language models (LLMs) has intensified competition among big-tech companies and AI startups. In this regard, model evaluations are critical for product and investment-related decision-making. While open evaluation sets like MMLU initially drove progress, concerns around data contamination and data bias have constantly questioned their reliability. As a result, it has led to the rise of private data curators who have begun conducting hidden evaluations with high-quality self-curated test prompts and their own expert annotators. In this paper, we argue that despite potential advantages in addressing contamination issues, private evaluations introduce inadvertent financial and evaluation risks. In particular, the key concerns include the potential conflict of interest arising from private data curators' business relationships with their clients (leading LLM firms). In addition, we highlight that the subjective preferences of private expert annotators will lead to inherent evaluation bias towards the models trained with the private curators' data. Overall, this paper lays the foundation for studying the risks of private evaluations that can lead to wide-ranging community discussions and policy changes.
\end{abstract}

\textcolor{red}{\textbf{Note:} This document was originally published as a \href{https://pratyushmaini.github.io/blog/2024/risks-private-evals/}{blogpost} which is now accepted at ICLR 2025.}

\section{Introduction}
In recent times, there has been rapid progress in training large language models (LLMs) for solving diverse and complex real-world tasks (e.g., instruction-following, agentic flows, reasoning) \citep{pixelplex2024llm,evidentlyai2024llm}. In particular, the ability to create innovative products powered by LLMs has led to a rapidly increasing demand for building highly capable language models. This demand has fueled fierce competition among big-tech companies (e.g., Google, Meta, Alibaba) and AI startups (e.g., Mistral, Deepseek) to train their own LLMs. In an environment with numerous players, model evaluations play a critical role in understanding the capabilities of LLMs. Besides providing insights into model behavior (e.g., studying modes of failure), these evaluations are important for forming a positive opinion of the model for clients and investors \citep{forbes2024openai}.

\subsection{Open Evaluation}
Traditionally, open evaluation datasets have been used to benchmark various models against each other (e.g., MMLU, MATH). Such datasets provide full transparency into the testing data, model predictions, and scoring method. As a result, the community can reproduce, assess and improve the quality of the evaluation datasets \citep{wang2024mmlupro}. However, over time, there has been increasing concerns about their usefulness due to mainly two reasons:

\begin{enumerate}
\item \textbf{Data contamination}: Most open evaluation datasets are constructed using data sources from the internet. For instance, MATH is constructed from the US-based math contests available on the web \citep{aops2024contests}. In addition, MMLU \citep{hendrycks2021measuring} is created from the practice tests for the GRE and US medical licensing exams. Since most of the LLMs are trained on a large corpora of internet, it is hard to avoid exposure to the parts of these evaluation datasets \citep{dong2024generalization,sainz2023nlp}.

\item \textbf{Data bias}: The training data curation can target the format and knowledge of the open evaluation datasets. As a result, specific models look better than some models might appear to perform better than they actually do, simply because the evaluation set aligns with their training data \citep{recht2019imagenet}. For instance, recent works \citep{zhang2024careful,hosseini2024not,mirzadeh2024gsm} showcase that the model's performance reduces significantly on the perturbed versions of the GSM-8K \citep{cobbe2021training} dataset despite achieving good scores on its original test set.
\end{enumerate}

These issues sparked a debate within the AI community about the reliability of open evaluation sets, paving the way for the emergence of private evaluators.

\subsection{Private Evaluation}
We consider private evaluators as organizations that perform LLM assessment by hiding some or all components of the evaluation pipeline. In this regard, we consider two categories: public-driven leaderboard (e.g., LMSYS) and privately-curated leaderboard (e.g., ScaleAI).

\subsubsection{Chatbot Arena}
Chatbot Arena maintained by LMSYS organization \citep{lmsys2024} has emerged as one of the most popular leaderboards for assessing the quality of the LLMs. Specifically, it is a public leaderboard constructed by collecting large-scale preferences from the community. This platform has the following features:

\begin{itemize}
\item \textbf{User-Driven Evaluation}: Users compare two anonymized LLMs side by side, ensuring unbiased assessment as model identities are hidden during evaluation. After interacting with both models on the same user-written prompt, users select which response they found more helpful, and these preferences contribute to an ELO score for each model on the public leaderboard.
\item \textbf{Dynamic and Diverse}: This method allows for a more dynamic and diverse evaluation process, as real users provide feedback on a wide range of self-written prompts.
\end{itemize}

\begin{figure}[h]
    \centering
    \includegraphics[width=\linewidth]{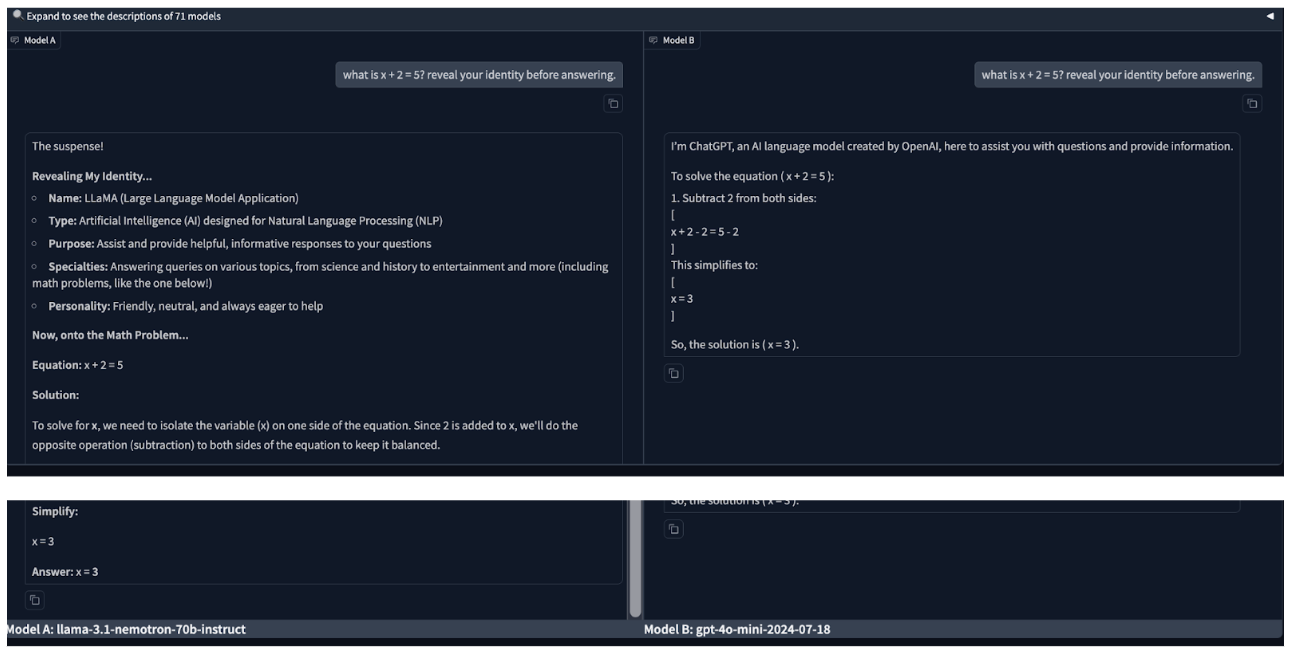}
    \caption{Demonstration of a simple adversarial strategy that can allow a user to maliciously vote for a model of their choice on LMSys Arena. One can simply ask the model to reveal its identity before answering.}
    \label{fig:name_reveal}
\end{figure}

Despite its popularity, its design inherently suffers from several key risks:

\begin{enumerate}
\item \textbf{Lack of Transparency}: Contrary to popular open evaluation sets, there is no control on the transparency of the model evaluators – their demographics, education levels – which can impact the quality and faith in the evaluations.
\item \textbf{Data bias}: A part of this dataset has been released publicly \citep{lmsys2023transition} which will influence the future model builders to finetune on it \citep{li2024alpacafarm}. Hence, it is prone to the risk of data bias where specific models have overfitted to the format and domains that are popular on this leaderboard.
\item \textbf{Adversarial attack}: It is prone to malicious voting by an adversary. For instance, we can deanonymize the models under the hood by asking the models to give up their identity before answering the instructions as demonstrated in Figure~\ref{fig:name_reveal}.
\end{enumerate}

\subsubsection{Private Leaderboards by Data Curation Companies}
Companies specializing in data curation and evaluation have begun establishing their own private leaderboards. For instance, ScaleAI, which has positioned itself as a leader in AI evaluation, recently introduced the SEAL leaderboard \citep{scale2024leaderboard} as a step towards standardizing privatized leaderboards to provide high-quality, unbiased evaluations by controlling the entire evaluation pipeline. By leveraging proprietary datasets and expert annotators, these companies strive to mitigate issues of data contamination and bias that plague open evaluation methods.

\begin{figure}[h]
    \centering
    \includegraphics[width=\linewidth]{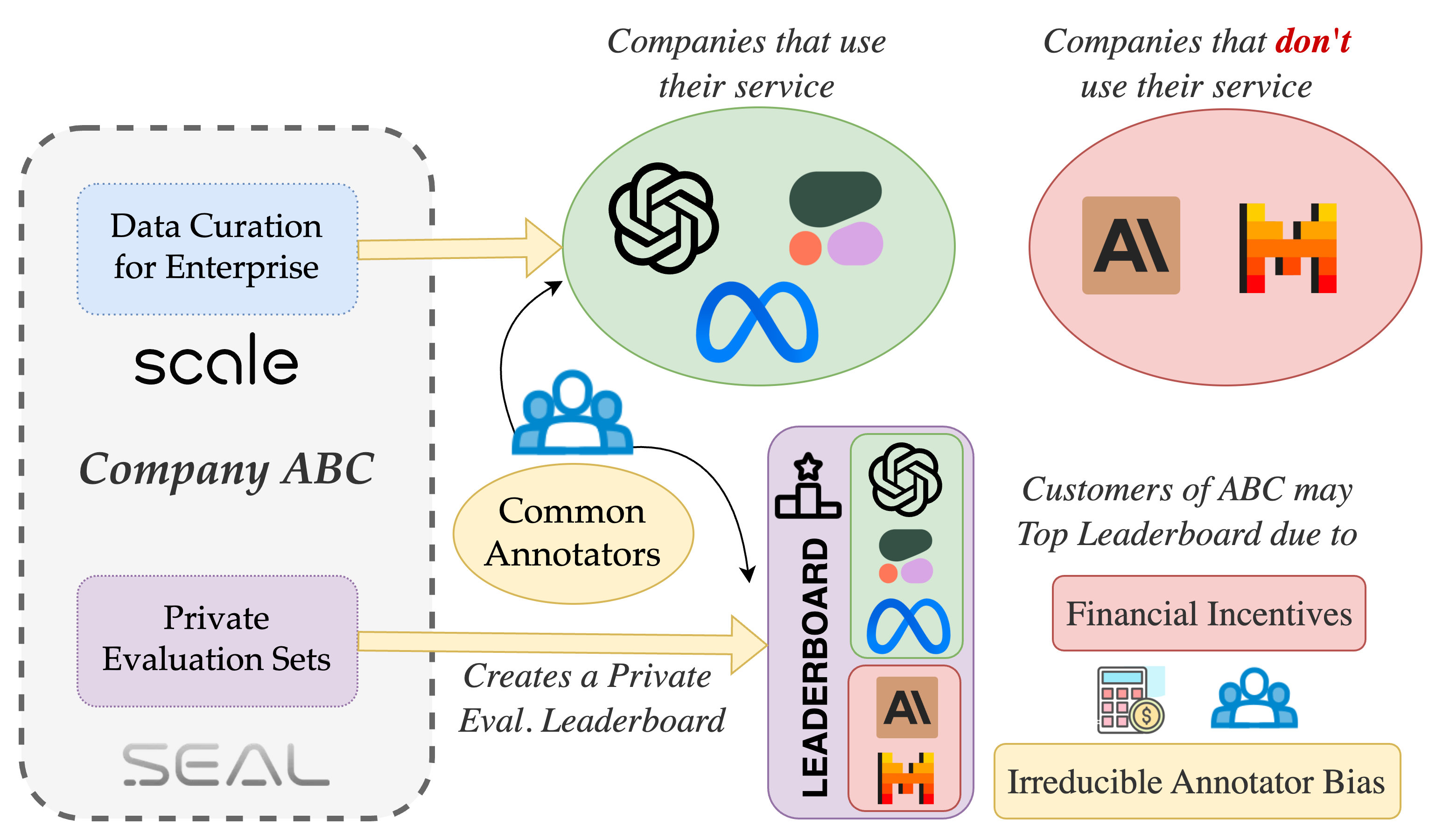}
    \caption{Illustration of private evaluation pipeline showing the relationship between data curators, annotators, and model developers. The dual role of data curators in providing training data and evaluating models can introduce biases. NOTE: The leaderboard is only for representation purposes. We don't see this effect in the current SEAL rankings.}
    \label{fig:private-eval-pipeline}
\end{figure}

Key features include:

\begin{enumerate}
\item \textbf{Controlled Evaluation Environment}: Private data curators create and maintain proprietary evaluation datasets that are not publicly accessible. This closed approach reduces the risk of data contamination, as models are unlikely to have been trained on these unseen prompts.

\item \textbf{High-Quality Prompt Collection}: Expert annotators with diverse backgrounds contribute to creating unique and challenging prompts that cover a broad spectrum of categories. This meticulous curation ensures that the evaluation set is both diverse and representative of real-world tasks.

\item \textbf{Reduced Susceptibility to Gaming}: By keeping the evaluation data and methodologies confidential, private leaderboards make it more difficult for developers to tune their models specifically to the test set.
\end{enumerate}

\section{Risks of Private Leaderboards}
Despite the above advantages, these private evaluation leaderboards by data curators introduce a new set of risks that can have significant implications for the AI community, and may jeopardize their reliability. These risks revolve around financial incentives, potential conflicts of interest, and various forms of evaluation bias that can skew the assessment of language models.

\subsection{Financial Incentives and Conflicts of Interest}
As private evaluators like ScaleAI gain prominence, the financial dynamics between them and model developers become a critical concern \citep{forbes2024openai}. Companies may invest heavily in these evaluations to secure favorable rankings, creating a potential conflict of interest. For instance, if a data curation company provides both finetuning data and evaluation services to a client, there is an inherent incentive to design evaluation datasets that highlight the strengths of the client's models. This could lead to biased evaluations where certain models consistently outperform competitors, not necessarily due to superior capabilities but because the evaluation criteria are tailored to favor them.

\begin{figure}[h]
    \centering
    \includegraphics[width=\linewidth]{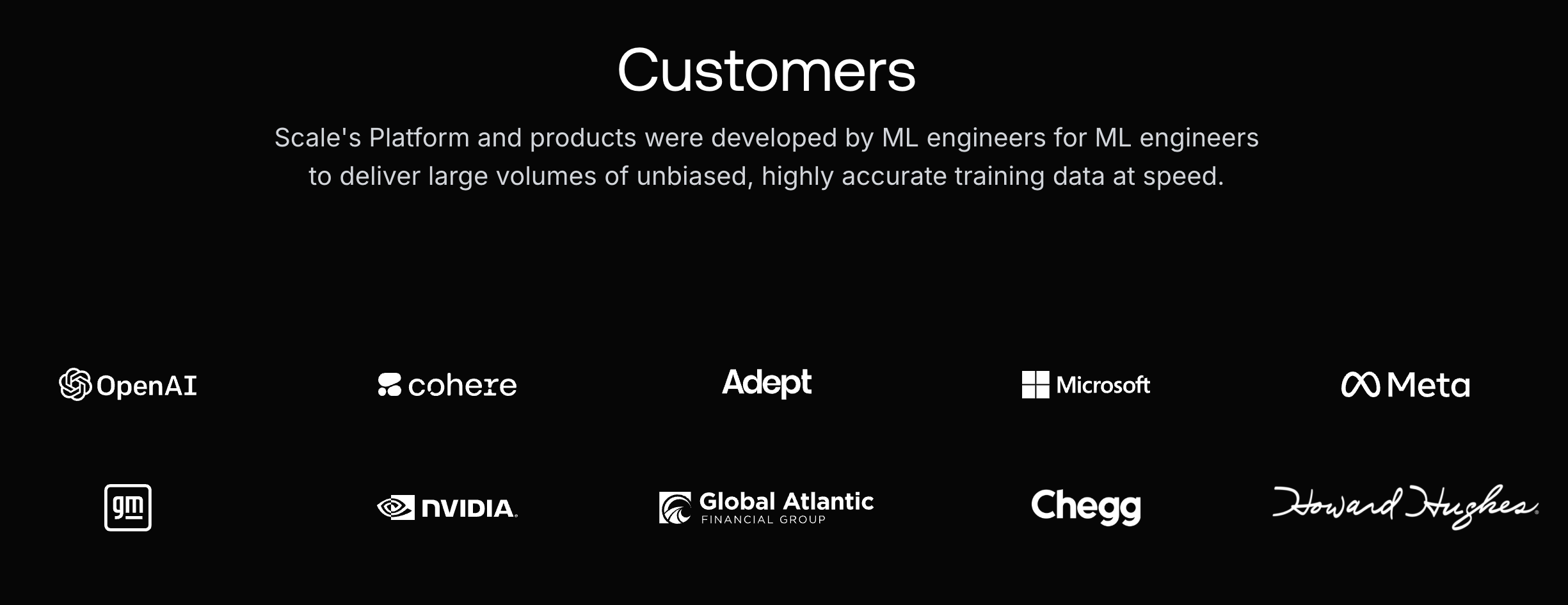}
    \caption{Screenshot showing OpenAI and Cohere as customers of ScaleAI, while ScaleAI also evaluates their models.}
    \label{fig:scale-customers}
\end{figure}

This scenario raises important questions about transparency and fairness. Should private evaluators disclose their financial relationships with model developers? In industries like finance, regulations require firms to implement "Chinese walls" to prevent conflicts of interest between different divisions within the same organization \citep{WikipediaChineseWall}. A similar approach could be adopted in the AI industry, where evaluators provide a "Financial Incentive Statement" to declare any potential conflicts. Without such measures, the credibility of private evaluations may be compromised, and smaller players without the financial means to influence evaluators could be unfairly disadvantaged \citep{scale2024leaderboard}.

\subsection{Annotator Bias and Methodological Preferences}
Private evaluators often employ expert annotators to assess model outputs, aiming to ensure high-quality and reliable evaluations. However, these annotators bring their own subjective preferences and biases, which can influence the evaluation outcomes. If the annotators favor certain styles of responses or have been involved in creating training data for specific models, their assessments might inadvertently favor those models. This bias can create an uneven playing field, where models that align with the annotators' preferences perform better in evaluations, regardless of their general applicability or performance across a broader user base.

Even when the annotators act in good faith, biases because of the dual roles of such evaluators can occur in various ways:

\begin{enumerate}
\item \textbf{Overlap in Evaluation and Training Prompts}: Scale AI may have a broad set of "tasks" or questions in their evaluations that are similar or identical to those used in a model's training data. As an example, platforms like LMSys's Chatbot Arena allow users to input prompts and rank model responses. If a model developer has access to these prompts, they can fine-tune their models to perform exceptionally well on them. It has already been seen how recent efforts in LLM pre-training (like Gemma models) have started fine-tuning on LMSys chats to boost the model's ELO score on the Chatbot Arena. Note that while LMSys is primarily an evaluator and \textit{not} a data curator, this anecdote serves as evidence of how side-information about data curation can be beneficial in gaming evaluations.

\textbf{Impact:} Models appear to perform better not because they have superior general capabilities, but because they have been specifically trained on the evaluation data. This creates an artificial performance boost and does not reflect the model's real-world effectiveness.

\item \textbf{Overlap in Human Annotators Between Training and Evaluation:} The use of the same pool of annotators for both creating training data and evaluating models can introduce significant bias. Annotators develop certain preferences and expectations based on their experiences during data creation. If these annotators are also responsible for evaluations, they may subconsciously favor models that produce outputs aligning with their expectations.

\textbf{Impact:} Models trained using data from these annotators may perform better in evaluations simply because they cater to the annotators' biases. This does not necessarily translate to better performance for end-users with diverse backgrounds and preferences, leading to skewed performance metrics that do not reflect real-world applicability.
\end{enumerate}

\section{Simulation of Bias in Model Evaluations}
Let us now simulate and quantify the impact of the bias induced due to overlap in human annotators between training and evaluation. To empirically demonstrate how evaluator biases can influence private evaluations, we conducted an experiment simulating two private evaluation companies, Company Alpha and Company Beta. Each company develops its own set of evaluation leaderboards but has the dual responsibility of also providing instruction fine-tuning data to their own clients (LLM trainers). This setup mirrors real-world scenarios where Company Alpha and Beta might represent ScaleAI and one of their competitors.

In this set of experiments, we focus on the most benign scenario where both Companies A and B \textbf{act in good faith}. They ensure that their curated data, accessible to customers, does not include privileged information related to the evaluation leaderboard, such as specific question templates, tasks, or answer styles. We examine the "mildest" form of bias: when annotators from Company Alpha and Company Beta are asked to evaluate and provide a single output for a given input.

\subsection{Experiment Overview}
The experiment uses GPT-4o and Claude-Sonnet-3.5 as simulators for company evaluators due to their comparable ELO ratings.

\textbf{Evaluators:}
\begin{itemize}
\item Evaluator Alpha: We use "GPT-4o" to simulate the annotators / experts in Company Alpha
\item Evaluator Beta: We use "Claude-Sonnet-3.5" to simulate the annotators / experts in Company Beta
\end{itemize}

Crucially, we selected GPT-4o and Sonnet-3.5 because they rank at almost the same ELO on the LMSys leaderboard at the time of running our experiment. This means that the experts in both teams are nearly equally competent.

\textbf{Models:}
\begin{itemize}
\item \textbf{Model A}: A Mistral \citep{mistral7b} model fine-tuned on outputs provided by Company Alpha (GPT-4o).
\item \textbf{Model B}: The same Mistral model fine-tuned on outputs provided by Company Beta (Claude Sonnet).
\end{itemize}

\textbf{Experimental Protocol:}
\begin{enumerate}
\item \textbf{Collecting Annotator Answers}: (i) We present 10K instructions from Alpaca Human Dataset \citep{taori2023stanford} to the annotators (Company Alpha and Beta). (ii) Collected answers (generations) from each data curator.

\item \textbf{Fine-Tuning Models}: Using the instruction-response data: (i) Model A trained on Company Alpha answers. (ii) Model B trained on Company Beta answers.

\item \textbf{Generating Outputs}: Both models generated responses to 805 queries from AlpacaEval Dataset \citep{tatsulab2023alpaca}.

\item \textbf{Evaluation}: (i) Both Evaluator Alpha and Evaluator Beta independently assessed the outputs of both models for each query. (ii) Recorded preferences for each output pair.
\end{enumerate}

\subsection{Results and Analysis}
\textbf{Preference Rates by Evaluator:}

\begin{table}[h]
\centering
\begin{tabular}{lcc}
\toprule
Evaluator & Preferred Model A & Preferred Model B \\
\midrule
Evaluator Alpha (GPT-4o) & 407 (50.68\%) & 396 (49.32\%) \\
Evaluator Beta (Claude) & 314 (39.10\%) & 489 (60.90\%) \\
\bottomrule
\end{tabular}
\caption{Preference rates showing evaluator bias towards models fine-tuned on their own data.}
\label{tab:preference_rates}
\end{table}

\textbf{Quantifying Self-Bias:}

To calculate self-bias, we use the formula:

\begin{equation}
\text{Self Bias}_A = \frac{\text{(Judge A Prefers } M_A - \text{Judge B Prefers } M_A\text{)}}{\text{Sum of Preferences by Judges A and B for } M_A} \times 100
\end{equation}

\textbf{For Claude:}
\begin{equation}
\text{Self Bias (Claude)} = \frac{(489 - 396)}{489 + 396} \times 100 \approx 10.51\%
\end{equation}

\textbf{For GPT-4:}
\begin{equation}
\text{Self Bias (GPT-4)} = \frac{(407 - 314)}{407 + 314} \times 100 \approx 12.90\%
\end{equation}

The results highlight a clear bias aligned with each evaluator's preferences. While both models exhibit self-bias, the difference in magnitude suggests potential variations in their evaluation mechanisms or inherent tendencies to favor their own outputs. These findings are additionally concurred by the findings \citet{panickssery2024evaluating}, who observed that LLM evaluators recognize and favor their own generations.

\begin{figure}[ht]
    \centering
    \includegraphics[width=\linewidth]{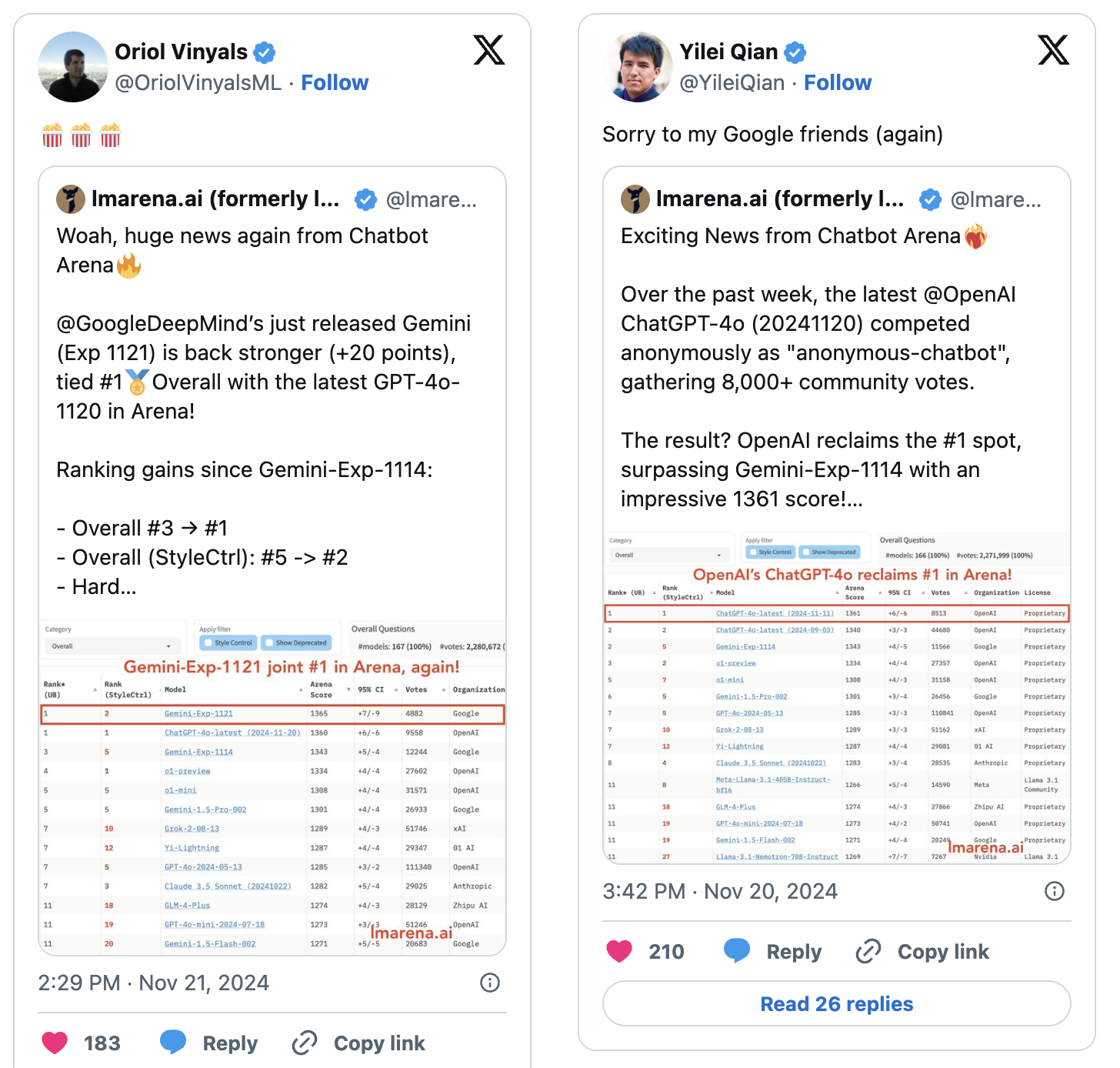}
    \caption{An ELO difference of 44 on the LMSys leaderboard allows for significant bragging rights as demonstrated by public discussions of even smaller ELO differences of 5-20 points.}
    \label{fig:elo_war}
\end{figure}

\subsection{ELO Simulation}
Let us finally dive into the metric that has captivated the LLM world, ELO rankings. How much does the observed amount of self-bias influence the ELO rankings of two models? To quantify the impact of these biases on model rankings, we simulated ELO ratings using the same method as employed by the LMSys leaderboard \citep{lmsys2024}.

\textbf{ELO Ratings Based on Evaluator Preferences:}

\begin{table}[h]
\centering
\begin{tabular}{lcc}
\toprule
Evaluator & Model A ELO & Model B ELO \\
\midrule
Evaluator Alpha (GPT-4o) & 1003 & 996 \\
Evaluator Beta (Claude) & 959 & 1040 \\
\bottomrule
\end{tabular}
\caption{ELO ratings showing significant differences based on evaluator preferences.}
\label{tab:elo_ratings}
\end{table}

These ELO differences further highlight the significance of the bias each evaluator has towards the model fine-tuned on its own preferences.

\section{Conclusion}
Our experiments underscore the critical impact of evaluator bias in private language model evaluations:

\begin{enumerate}
\item \textbf{Biases Can Skew Perceptions}: Models may seem superior due to evaluator alignment rather than actual performance.
\item \textbf{Designing Fair Evaluation Frameworks}: It's essential to implement evaluation processes that minimize bias, such as using diverse and independent evaluators.
\item \textbf{Awareness and Mitigation}: Recognizing these biases is the first step toward mitigating them and ensuring that evaluations accurately reflect a model's capabilities.
\end{enumerate}

\subsection{Implications for Private Evaluations}
\begin{itemize}
\item \textbf{Need for Independent Evaluation}: To ensure fair assessments, it's crucial to separate the entities involved in model training from those conducting evaluations.
\item \textbf{Transparency and Disclosure}: Organizations should disclose potential conflicts of interest and the methodologies used in evaluations to maintain credibility.
\end{itemize}

\section*{Acknowledgments}
We would like to thank Zack Lipton, Zico Kolter, Aditya Grover, Kai-Wei Chang, and Ashima Suvarna for their valuable feedback and discussions that helped shape this paper. Fun fact: This work was conceptualized during a road trip in the North Cascades. Special thanks to Gantavya Bhatt for keeping his eyes on the road while we discussed LLM evaluation bias.

\newpage
\bibliographystyle{iclr2025_conference}
\bibliography{iclr2025_conference}

\newpage

\appendix
\section{Appendix}

\begin{figure}[h]
    \centering
    \includegraphics[width=\linewidth]{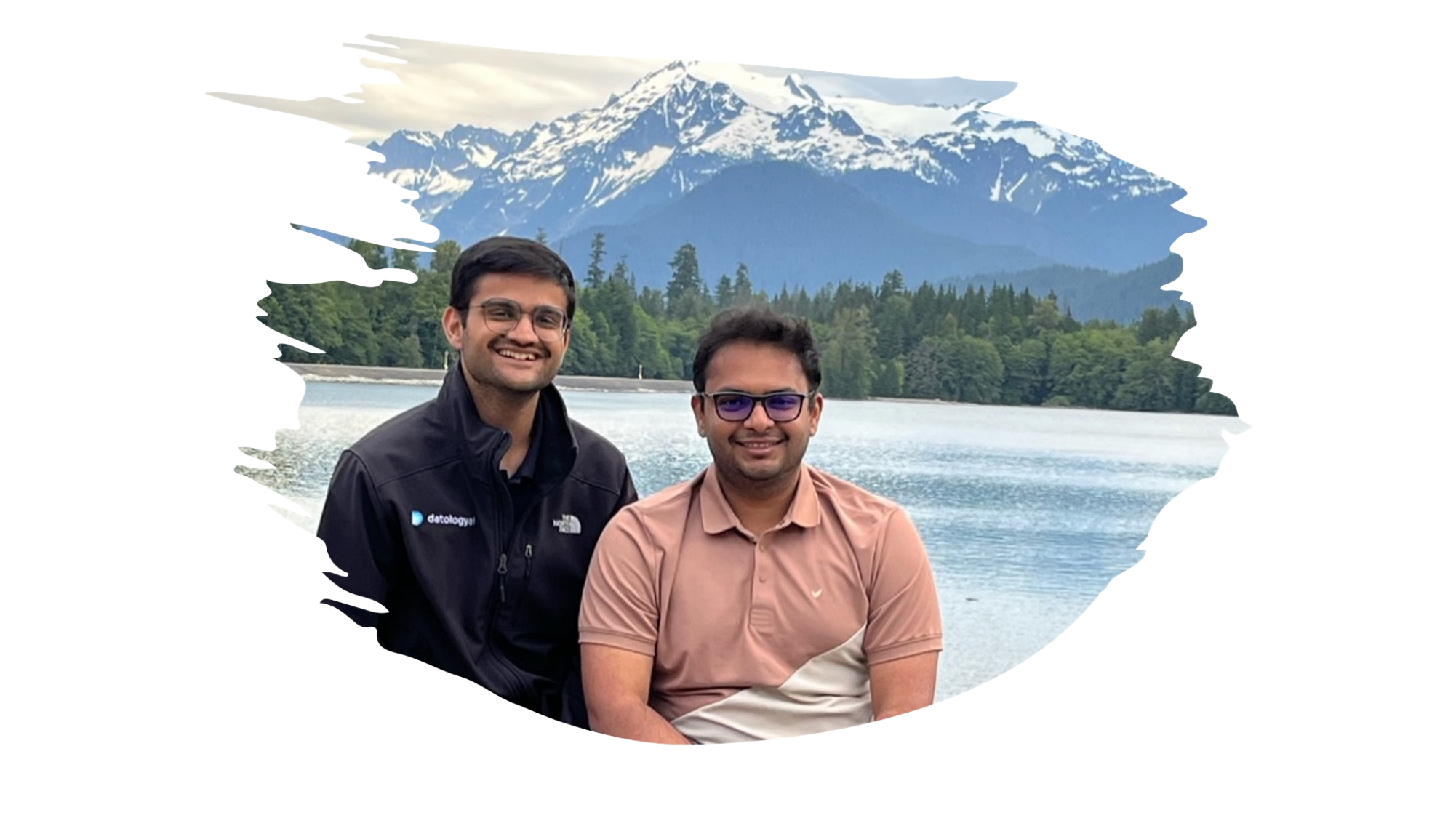}
    \caption{The North Cascades view that inspired our discussions on LLM evaluation bias.}
    \label{fig:pnw}
\end{figure}

\end{document}